# Effect of ball milling and post annealing on structural and magnetic properties in $Ni_{50}Mn_{36}Fe_2Sb_{12}$ Heusler alloy

Roshnee Sahoo[1,4*], K. G. Suresh[1], A. K. Nigam[2], X. Chen[3] and R. V. Ramanujan[3]

[1]Department of Physics, Indian Institute of Technology Bombay, Mumbai 400076, India

[2]Tata Institute of Fundamental Research, Mumbai 400005, India

[3]School of Materials Science and Engineering, Nanyang Technological University, Singapore 639798

[4] Max Planck Institute for Chemical Physics of Solids, 01187 Dresden, Germany

**Abstract:**

The effect of ball milling on the structural, magnetic and exchange bias properties of $Ni_{50}Mn_{36}Fe_2Sb_{12}$ Heusler alloys was studied. The ball milled samples exhibited coexisting austenite and martensite phases at room temperature, while annealing supresses the austenite phase completely. Ball milling was found to reduce the grain size, which resulted in the weakening of the ferromagnetic properties. An exchange bias field of 111 Oe and coercivity of 826 Oe were observed at 5 K in the as-milled sample, in contrast to the bulk alloy values of 288 Oe and 292 Oe, respectively. Annealing causes an increase in the ferromagnetic ordering and a decrease in the interfacial exchange coupling, resulting in a decrease of both exchange bias and coercivity.

**Keywords:** A. Shape memory alloys, B. Annealing, C. Magnetic Properties

Recently Ni-based full Heusler alloys have been receiving considerable attention due to their anomalous magnetic properties, which are associated with magneto-structural phase transition. Phenomena such as the shape memory effect[1,2], magnetocaloric effect (MCE)[3,4], magnetoresistance (MR)[5,6] and exchange bias (EB)[7,8] are directly influenced by the strong magneto-structural coupling observed in these alloys. This first order magneto-structural transition (martensitic transition) results in a phase transition from an austenite phase (cubic) to a martensite phase (tetragonal or orthorhombic) on cooling.[9-11] The magnetic properties of these alloys can be mainly attributed to the Mn magnetic moment.[12] The Mn-Mn exchange coupling is ferromagnetic (FM) when Mn occupies the usual Mn site and it is antiferromagnetic (AFM) when Mn occupies both Mn and Sb sites.[13] This is because the Mn-Mn interatomic distance plays a crucial role in determining the nature of exchange interactions.[14] The complex magnetic state of the martensite phase is responsible for phenomena like the exchange bias (EB) effect in these alloys.[15,16]

Ball milling is a non-equilibrium method which produces stable/metastable states, e. g., amorphous and nanostructured alloys. This method has attracted much attention because of its applications in the synthesis of applied materials in various fields. As the system changes from bulk to fine particles/ball milled, the associated magnetic properties usually change considerably. However, from the Heusler family, very few studies have been reported of the effects of ball milling. One of these studies is related to investigations of Ni-Mn-Ga nanoparticles prepared by ball milling.[17,18] It was found that with increasing milling time, the structural transition gradually disappears. Similarly, in Ni-Mn-Sn alloys short time milling suppresses the ferromagnetic (FM) state considerably.[19] Ball milling reduces grain size and increases chemical disorder and mechanical strain in the alloy. Hence, the magnetic properties as well as structural transition are affected. Exchange bias behavior and other magnetic properties of bulk Fe doped $Ni_{50}Mn_{38}Sb_{12}$ Heusler alloy, have been extensively studied very recently.[20] The results obtained in bulk $Ni_{50}Mn_{36}Fe_2Sb_{12}$ revealed that the martensitic transition occurs at 274 K and the magnetization was 32 emu/g at 290 K. This alloy shows the EB effect, with a maximum EB field of 288 Oe at 5 K for a cooling field of 50 kOe. In order to study the effect of grain size reduction and modification on EB, we have studied the magnetic properties of fine particles of this alloy obtained by ball milling. In this paper, we present a comparative study of the magnetic properties and the EB effect of as-milled and annealed powders of $Ni_{50}Mn_{36}Fe_2Sb_{12}$ alloy with those of the bulk alloy.

Polycrystalline samples of $Ni_{50}Mn_{36}Fe_2Sb_{12}$ were prepared in an argon atmosphere by the arc melting process. The constituent elements were of at least 99.99% purity. The ingot was remelted several times and the weight loss after the final melting was found to be less than 0.2%. For homogenization, the as-cast sample was sealed in evacuated quartz tube and subsequently annealed for 24 hr at 850 ˚C. The annealed sample was ball milled with 200 rpm in $N_2$ atmosphere for 15 hr using tungsten carbide balls. Two batches of the ball milled sample were annealed at two different temperatures, i.e, 500 °C and 800 °C, in vacuum for 2 hr. The field emission gun-scanning electron microscopy (FEG-SEM) measurements have been carried out using a JSM-7600F SEM. Structural characterization was performed from the powder x-ray diffraction (XRD) using Cu-Kα radiation. Magnetization measurements were carried out using a vibrating sample magnetometer attached to a Physical Property Measurement System (Quantum Design, PPMS-6500).

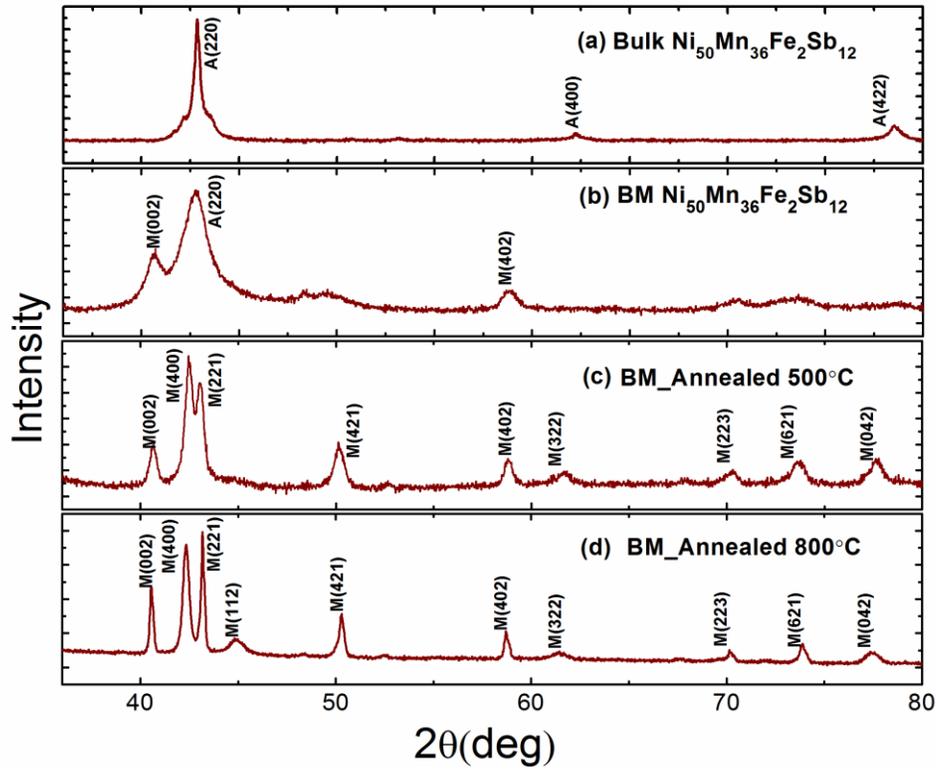

FIG. 1. X-ray diffraction patterns of $Ni_{50}Mn_{36}Fe_2Sb_{12}$ (a) bulk, (b) as milled, (c) annealed at 500°C for 2 hr, (d) annealed at 800°C for 2 hr.

Figure 1 shows the x-ray diffraction patterns of as-milled as well as annealed powders of $Ni_{50}Mn_{36}Fe_2Sb_{12}$. For the bulk alloy, $L2_1$ austenite phase along with a small fraction of martensite phase was observed at room temperature (figure 1(a)). For the ball milled sample, the sample still retains long range order, as confirmed by well defined peaks in figure 1(b). By comapring the bulk and the as-milled samples, it can be seen that the as-milled sample has both the austenite and the martensite phases at room temperature. However, after annealing, the XRD pattern [figure 1(c) and (d)] reveals the orthorhombic martensitic phase, the peaks corresponding to the austenite phase have disappeared. As expected, the XRD peaks of the as-milled sample are broad, which become sharp upon annealing. Therefore, it is clear that ball milling followed by annealing results in a completely martensitic structure in this alloy, while the bulk form has a purely austenite stucture. This difference is expected to play a crucial role in their magnetic properties.

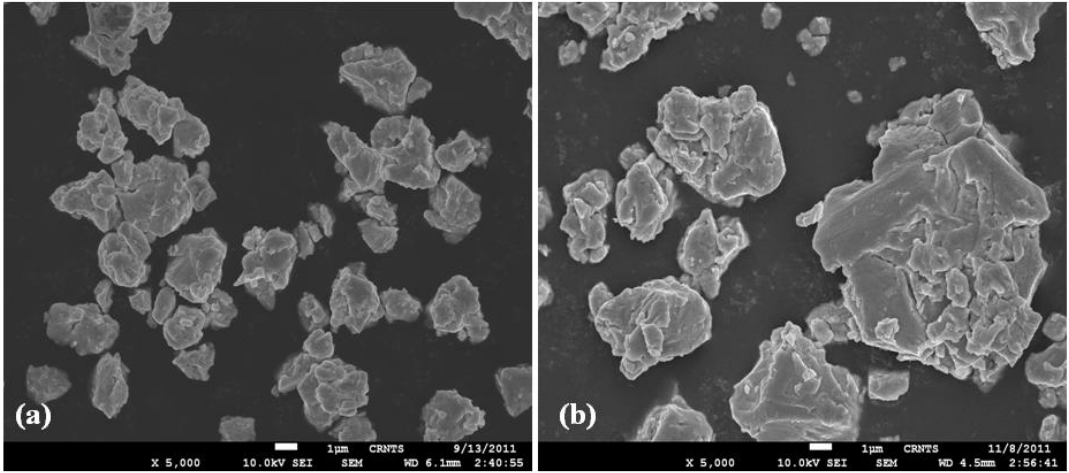

FIG. 2. FEG−SEM images of $Ni_{50}Mn_{36}Fe_2Sb_{12}$ (a) as milled, (b) annealed at 800°C for 2 hr of $Ni_{50}Mn_{36}Fe_2Sb_{12}$ ball milled sample.

From the FEG-SEM micrographs (figure 2), it is found that the paricle size is in the range of 0.5-5 µm for the as-milled sample. After annealing at 800°C, the range of particle size is found to be 1-15 µm. The broadening of peaks as a result of ball milling is mainly due to strain energy that arises due to the decrease in grain size. In FeB and FeCu ball milled systems, it has been reported that the internal strain increases with increase in milling time.[21,22] From the Williamson-Hall equation, it has been found that the functional dependence between grain size and atomic level strain is inversely proportional.[23,24] Annealing increases grain size, which results in a decrease in the width of XRD peaks. Annealing process also aggravates the lattice oscillation. This may help to get rid of any lattice dislocation, which consequently reduces the lattice strain energy.

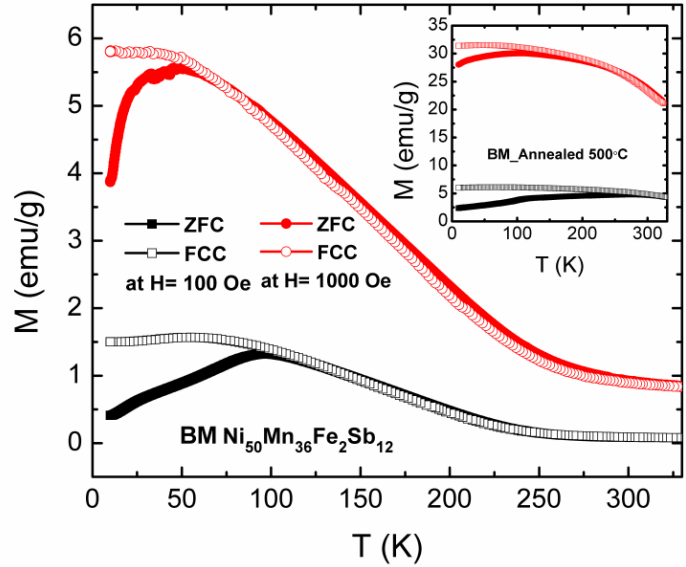

FIG. 3. Temperature variation of magnetization in zero field cooled (ZFC) and field cooled cooling (FCC) modes for as-milled $Ni_{50}Mn_{36}Fe_2Sb_{12}$ in 100 Oe and 1000 Oe field. The inset shows the same curves for the sample annealed at 500°C.

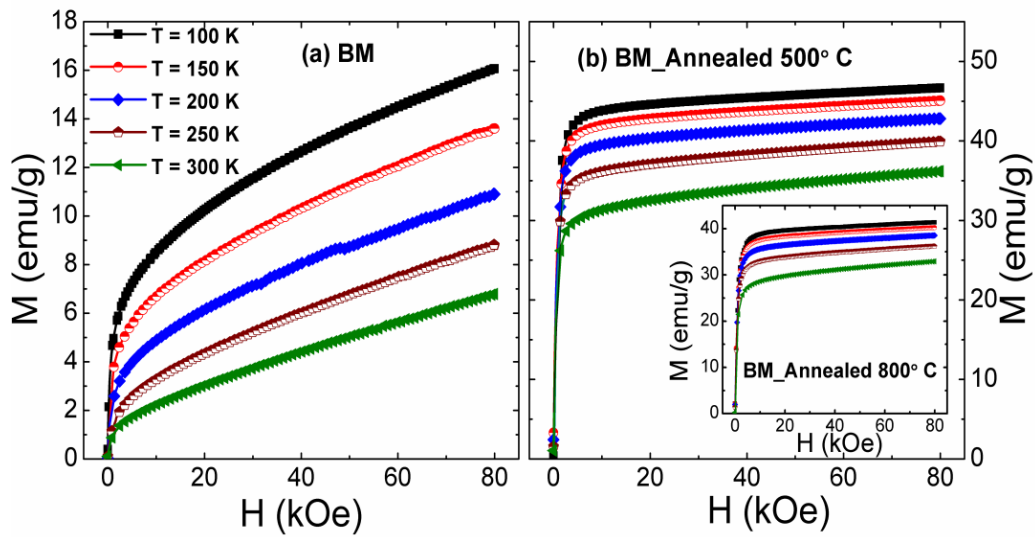

FIG. 4. Magnetization isotherms of $Ni_{50}Mn_{36}Fe_2Sb_{12}$ (a) as-milled (BM) and (b) annealed at 500°C. The inset shows the data for the sample annealed at 800°C.

The temperature variation of magnetization for the ball milled sample is shown in figure 3, the inset shows the corresponding plots for the sample annealed at 500 °C, for two different fields. The martensitic transition (at $T_M$=275.5 K) reported in the bulk alloy is absent in both the cases.[20] A similar observation has been made in the Ni-Mn-Ga and Ni-Mn-Sb systems, where a reduction in grain size suppresses the structural transition.[25,26] However, ball milled Ni-Mn-Ga alloys show martensitic transition after annealing.[27] Furthermore, in the present case, compared to the bulk system, the magnetic moment is lower for the ball milled system, which is attributed to a decrease in the strength of ferromagnetic exchange interactions due to the small particle size. It can be seen from the inset of figure 3 that in the case of the annealed sample, the magnetic moment has increased considerably. It can also be seen that thermomagnetic irreversibility is present at both in 100 Oe and 1 kOe. The clear difference in ZFC and FCC curves (figure 3) in the as-milled and annealed powders indicates that the magnetic state is inhomogenous. This can be ascribed to spin glass like behavior and probably arises due to chemical/atomic disorder upon ball milling.[28] Figure 4 shows the magnetization curves for a maximum field of 70 kOe in the temperature range of 100 - 300 K. The absence of saturation trend in the as-milled sample changes to near saturation behavior after annealing. The magnetization value at high fields in the annealed samples is comparable to that of the bulk.

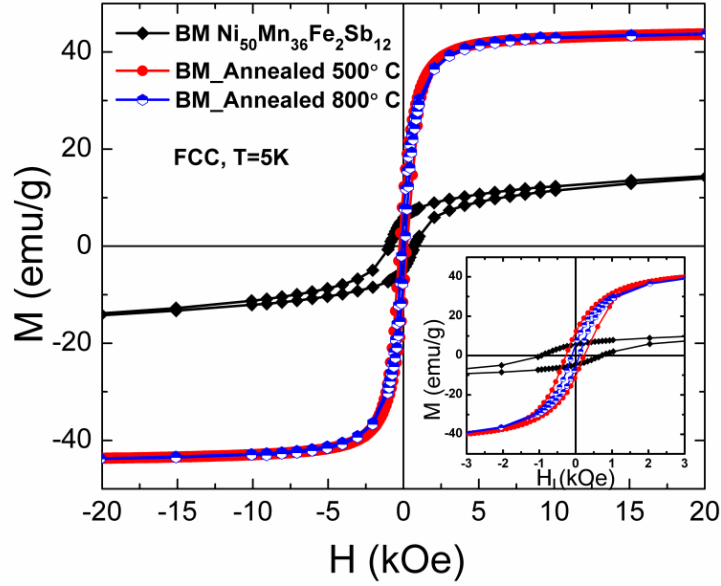

FIG. 5. Magnetic hysteresis loop at 5 K after field cooling in 50 kOe for the as-milled and annealed (500°C and 800 °C) samples of $Ni_{50}Mn_{36}Fe_2Sb_{12}$. The inset shows the low field portion of the plots shown in the main panel.

As shown in figure 5, the exchange bias properties of the ball milled samples were investigated by measuring the field cooled (FC) hysteresis loop at 5 K. During field cooling, 50 kOe field was applied to the sample at 300 K which was subsequently cooled down to 5 K. Once the measuring temperature was reached, the hysteresis loop was recorded for a maximum field of 20 kOe. For the sake of clarity, we show in the inset the hysteresis loop upto 3 kOe. The shift in the curve towards the negative field direction, defined as exchange bias, is clearly seen for the as-milled sample. EB field is quantified as $H_{EB}= -(H_1+H_2)/2$, the coercivity is defined as $H_C= (H_1-H_2)/2$. Here $H_1$ and $H_2$ are the points where magnetization curve intersects the field axis in the positive and negative field axis respectively. Table 1 shows the variations of $H_{EB}$ and the $H_C$ as a function of annealing temperature. For the as-milled sample, maximum $H_{EB}$ is found to be 111 Oe. Increase in atomic disorder and strain result in the reduction in unidirectional anisotropy after ball milling. Upon annealing at 500 °C, exchange bias further reduces to 17 Oe and completely vanishes when annealed at 800 °C. This is attributed to the increase in the grain size (as evidenced by the powder x-ray diffraction data and FEG-SEM data), which causes enhancement in FM coupling (as evidenced by the M-T and M-H data) and a decrease in the FM-AFM interfacial coupling. Generally in fine particles, the EB effect tends to vanish with increase in the FM component, due to the decrease in the surface to volume ratio brought about by annealing.[29] It is also important to note that the bulk alloy of the same composition shows an exchange bias field of 288 Oe under identical experimental conditions.[20] This implies that though the saturation magnetization is almost restored by annealing, the local microstructure and magnetic structure in the annealed samples are different from those of the bulk.

Table. 1 Variation of exchange bias field and coercivity in bulk, as-milled and annealed samples of $Ni_{50}Mn_{36}Fe_2Sb_{12}$ at 5 K, after field cooling in a field of 50 kOe.

| Sample | $H_{EB}$ (Oe) | $H_C$(Oe) |
| --- | --- | --- |
| As-milled | 111 | 826 |
| Annealed at 500 °C | 17 | 250 |
| Annealed at 800 °C | 0 | 102 |
| *Bulk* | *288* | *292* |

It has also been observed that the coercivity value of 826 Oe for the as-milled sample reduces to 250 Oe and 102 Oe, after annealing at 500 ºC and 800 ºC, respectively. In the bulk form, the coercivity was found to be 292 Oe.[15] Development of large magnetic anisotropy due to decreasing grain size and the generation of strain as a result of ball milling gives rise to increase in the coercivity in the as-milled sample. When particle size is reduced, micro strains are introduced in the FM phase.[21,22] Such a high magnetic anisotropy has also been reported in ball milled alloys of $PrCo_3$ and $GdAl_2$.[30,31] The decrease in the coercivity with annealing is due to the fact that the internal strain and the disorder are gradually reduced, resulting in a reduced strain-induced magnetic anisotropy.

At this point, it is of interest to compare the present results with those obtained in $Ni_{45}Co_5Mn_{38}Sb_{12}$ alloy, which showed significant enhancement of EB field and coercivity upon ball milling.[23] The sharp martensitic transition seen in the bulk $Ni_{45}Co_5Mn_{38}Sb_{12}$ alloy was suppressed to a very weak transition upon ball milling. Both the EB field and the coercivity were found to increase considerably after ball milling, which was attributed to the presence of soft and hard magnetic phases resulting from a wide range of particle size distribution. Therefore, the complete absence of the martensitic transition in the present case is one of the reasons for the nominal change in the EB field and the coercivity. It may be noted that while the particle size distribution in ball milled $Ni_{45}Co_5Mn_{38}Sb_{12}$ alloy is in the range of 1-50 μm, it is in the range of 0.5-5 μm in the present case. This difference may also contribute to smaller changes in EB field and the coercivity in the present case.

In summary, as-milled and annealed powders of $Ni_{50}Mn_{36}Fe_2Sb_{12}$ were investigated by XRD and magnetometry. The broadening of XRD peaks of as-milled sample is attributed to the reduction in grain size, which is also observed from the FEG-SEM analysis. Effect of strain must also be a contributing factor to the broadening. Upon ball milling, the martensite phase appears to gain strength, which gets further enhanced upon annealing. This is also evidenced by the absence of the martensitic transition in the as-milled and the annealed powders. Exchange bias decreases upon ball milling and subsequent annealing, which is attributed to decrease in exchange coupling between FM and AFM phases as a result of grain size reduction and atomic disorder. By comparing with the bulk alloy, it is found that ball milling increases coercivity as a result of mechanical strain and the disorder induced during the ball milling. Annealing is found to reduce the coercivity of the milled sample.

RVR acknowledges the support of NTU-HUJ-BGU Nanomaterials for Energy and Water Management Programme under the Campus for Research Excellence and Technological Enterprise (CREATE), that is supported by the National Research Foundation, Prime Minister's Office, Singapore and the support of U.S, AOARD, Tokyo (Dr. K. Caster, program manager). KGS thanks DST, Govt. of India for supporting this work and CRNTS, IIT Bombay for FEG-SEM facility.